\documentclass[[aps,prb,twocolumn,showpacs,preprintnumbers,amsmath,amssymb]{revtex4}
\usepackage{graphicx}
\usepackage{epstopdf}
\usepackage{xcolor}

\begin{document}

\title{Biocompatible technique for nanoscale magnetic field sensing with Nitrogen-Vacancy centers}

\author{E. Bernardi$^1$, E. Moreva$^1$, P. Traina$^1$, G. Petrini $^{2,1}$, S. Ditalia Tchernij$^2$, J. Forneris$^{2,3}$, \v{Z}. Pastuovi\'{c}$^4$,  I. P. Degiovanni$^{1,3}$,   P. Olivero$^{2,3,1}$, M. Genovese$^{1,3}$.}
\affiliation{$^1$Istituto Nazionale di Ricerca Metrologica, Strada delle cacce 91, Torino, Italy}
\affiliation{$^2$Physics Department and NIS Centre of Excellence - University of Torino, Torino, Italy}
\affiliation{$^3$Istituto Nazionale di Fisica Nucleare (INFN) Sez. Torino, Torino, Italy}
\affiliation{$^4$Centre for Accelerator Science, Australian Nuclear Science and Technology Organisation, New Illawarra rd., Lucas Heights, NSW 2234, Australia}

\begin{abstract}
We present an innovative experimental set-up that uses Nitrogen-Vacancy centres in diamonds to measure magnetic fields with the sensitivity of $\eta=59.6\pm1.3 \, \textrm{nT}/\sqrt{\textrm{Hz}}$ at demonstrated nanoscale. The presented method of magnetic sensing utilizing the ODMR technique for the optical detection of microwave-driven spin resonances induced in NV centres, which are highly concentrated to a nanoscaled volume probed by focused laser light, is characterized by the excellent magnetic sensitivity at such small scale and the full biocompatibility. As such, this method offers a whole range of research possibilities for biosciences. We also report how the magnetic sensitivity changes for different applied laser power and discuss the limits of the sensitivity sustainable with biosystem at such small volume scale.

\end{abstract}

\newcommand{\ket}[1]{\mbox{\ensuremath{|#1\rangle}}}
\newcommand{\bra}[1]{\mbox{\ensuremath{\langle#1|}}}
\flushbottom
\maketitle
\thispagestyle{empty}

Magnetometry in biological systems is of the utmost importance for fundamental biological science and
medicine.
Mapping brain activity by recording magnetic fields produced by the electrical currents which are naturally occurring 
in the brain is of extreme interest\cite{xiong2003directly,li2017lifting}, with direct applications in the timely detection and cure of psychic and neurodegenerative disorders \cite{kullmann2010neurological,waxman2006axonal, tomagra2019quantal}. Measuring the magnetic fields produced by electrical currents in the heart is also of the utmost importance\cite{cohen1967magnetic}, since this also could lead to a new generation of non-invasive diagnostic and therapeutic techniques \cite{wacker2014diagnosis}. Superconductive quantum interference device (SQUID) magnetometers are usually used for both these kinds of measurements. Nonetheless, the significant drawbacks of SQUID magnetometers are represented by the facts that they are unable to sense to single nerve impulses, and they are costly, bulky, and require cryogenic refrigeration \cite{faley2006new, baudenbacher2003monolithic}.

Magnetometers based on Nitrogen-Vacancy(NV) centers in diamond represent a valid alternative to SQUID-based magnetometry. Firstly, diamond offers the substantial advantage of being fully biocompatible \cite{tomagra2019quantal, yu2005bright, zhu2012biocompatibility, guarina2018nanodiamonds, schrand2007diamond, liu2007biocompatible}. On the other side, NV centers are characterized by a peculiar electronic level structure
that allows the optical detection of their microwave-driven
spin resonances with a technique referred as Optically Detected Magnetic
Resonance(ODMR)\cite{gruber1997scanning,neumann2010single,robledo2011high,childress2006coherent}. The shift in the ODMR frequency is related to the projection of the local magnetic field along the NV center axis. If an ensemble of NV cen-ters is used, it is possible to reconstruct the 3D structure
of the field, taking advantage of the 4-different possible orientations of the NV axis within the surrounding crystal structure.\cite{rondin2014magnetometry}.

NV magnetometry has already been exploited to detect the action potential in a macroscopic biological sample\cite{barry2016optical}. In this proof-of-principle experiment, Barry and coworkers achieved a magnetic field sensitivity of $\eta=15\pm1\, \textrm{pT}/\sqrt{\textrm{Hz}}$ with a sensing volume, i.e.the volume containing the ensemble of the excited centers, of $(13 \times 2000 \times 2000)\,\mu\textrm{m}^3$ and optical power of 2.75 - 4.5 W. To extend NV-based-biomagnetometry from macroscopic samples to small tissues up to single cells, a smaller sensing volume and a lower optical power should be used. The characteristic size of a cell is approximately $10\,\mu\,\textrm{m}$, and the laser power used in confocal measurement on living cells reported in \cite{mcnally2005comparative} is
in the range of few mW.

In this work, we present a NV-magnetometry protocol characterized by a sensing volume of $(0.01 \times 10 \times 10)\,\mu\textrm{m}^3$ and optical powers from 2.5 mW to 80 mW, values compatible, as we will discuss, with magnetometry at the cellular level.

\section{Experimental Setup}

\begin{figure}[!h]
\centering
\includegraphics[scale=0.25]{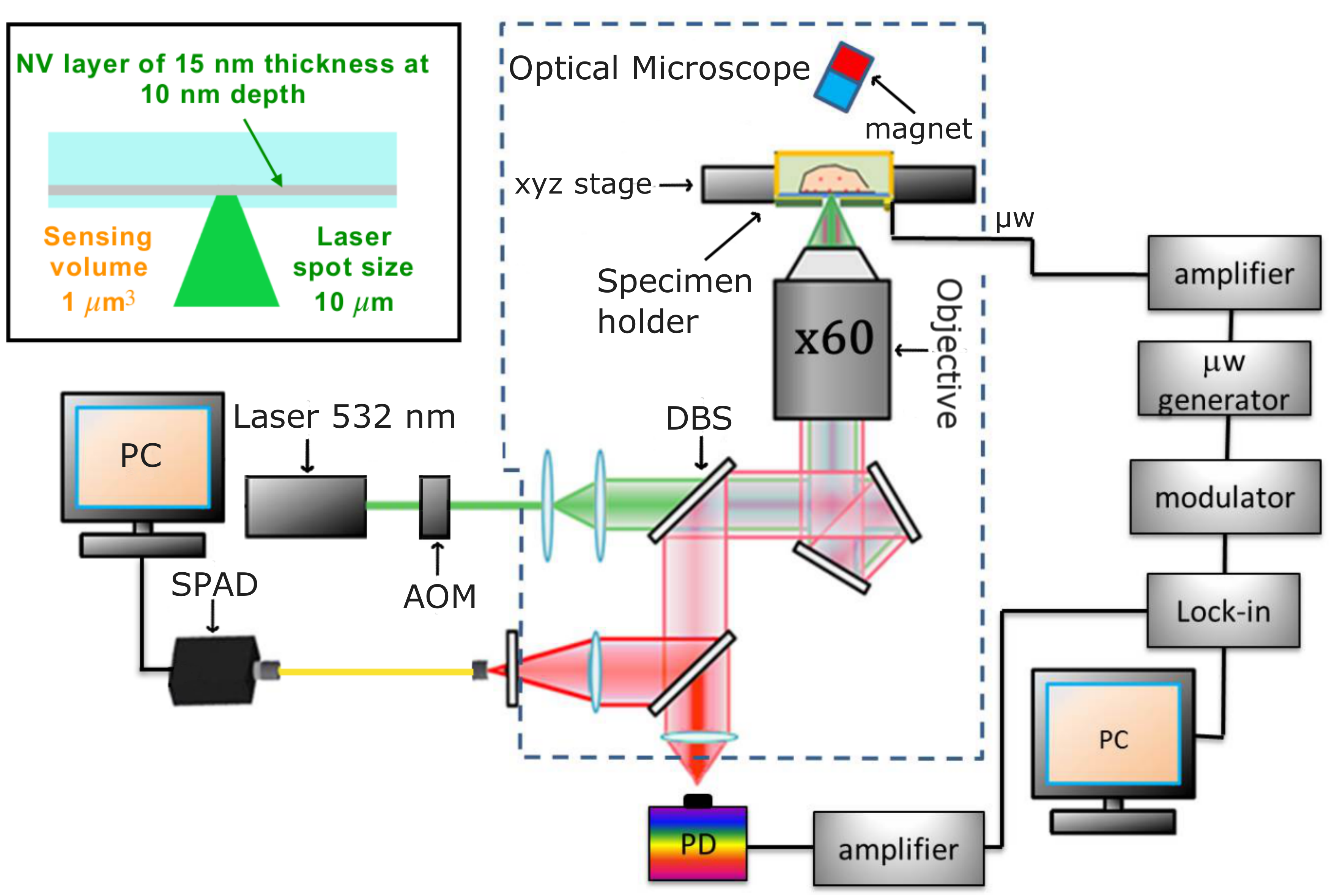}
\caption{Schematics of the experimental set-up. The optical excitation using a green laser, the microwave($\mu$w) control and the lock-in detection are depicted. In the inset a drawing of the sample is shown.}
\label{fig:set_up}
\end{figure}

The application of a magnetic field removes the energy degeneracy between the $m_S=\pm1$ spin states, and the frequency difference
$\nu_+-\nu_-$ between the two ODMR dips is
proportional to the component $B_{NV}$ of the field along
the NV-axis:
\begin{equation}
\nu_+-\nu_-=\frac{1}{h}2g\mu_BB_{NV}
\label{eq:BNV} 
\end{equation}
where $g$ is the Land\'e factor and $\mu_B$ the Bohr magneton. 
A variation $\delta B_{NV}$ in the applied magnetic field causes a shift $\delta\nu_+=\frac{1}{h}g\mu_B\delta B_{NV}$ to the higher-frequency ODMR
dip and a corresponding $\delta\nu_-=-\frac{1}{h}g\mu_B\delta B_{NV}$  shift for the lower-frequency
one. Tracking the ODMR shift $\delta\nu_+$ allows the measurement of the variation of the applied field $\delta B_{NV}$.

The simplest way to track the ODMR shift is to collect the photoluminscence signal while scanning the microwave frequency. Adopting a frequency modulation of the microwaves can improve this method: the modulating signal is centered at the resonance dip and has an amplitude equal to the full-width half maximum of the resonance \cite{schoenfeld2011real}. The resulting modulated photoluminescence signal is read by a lock-in amplifier(LIA).

Figure  \ref{fig:set_up} depicts the experimental set-up: the diamond sensor, laser excitation system, the microwave generation and LIA detection apparatus. 
The diamond sample was mounted on a microwave planar ring antenna, specifically designed for ODMR measurements in a 400 MHz frequencies range centered around the 2.87 GHz spin resonance\cite{sasaki2016broadband}.

The sensor consisted of a $3\times3\times0.3$ mm$^{3}$ diamond substrate produced by Element Six by CVD deposition, having $<1$ ppm and $<0.05$ ppm concentrations of substitutional nitrogen and boron, respectively. The substrate incorporates  a $\sim 10$ nm thick layer of NV centers at a concentration of $n_{NV} \sim 3 \cdot 10^{19}$ cm$^{-3}$, see inset of Fig. \ref{fig:set_up}, that was produced by low-energy N ion irradiation
followed by high temperature annealing.

The excitation light (80 mW optical power) at 532 nm was obtained by the second harmonic of a Nd:YAG laser with
high power stability (Coherent Prometheus 100NE) and was focused close to the bottom surface (i.e. the one incorporating the NV layer) of the diamond sample through an air objective (Olympus UPLANFL) with
Numerical Aperture NA=0.67. The spot size of the focused laser beam is $\sim(10 \times 10 )\,\mu\textrm{m}^2$. The power of the excitation light is varied using a Neutral Density filter. An Acousto Optic Modulator (not shown in Fig. 1) after the laser source is exploited to switch on and off the laser illumination on the sample. This solution allows to shine the laser on the sample only during the measurement time, reducing the total amount of light energy delivered to the sample. This is of key importance in biological applications.  

The microwave control was obtained by a commercial microwave  generator (Keysight N5172B) whose central frequency was internally modulated at $f_{mod} = 1009$ Hz with modulation depth $f_{dev}=0.6$ MHz. For simultaneous hyperfine driving, the microwave was mixed via a double-balanced mixer with a $\sim$ 2.16 MHz sinewave to create two simultaneous driving modulated frequencies near the central frequency. The microwave generator  was connected to the lock-in
LIA to provide a sinusoidal reference channel modulated at $f_{mod}$.

A permanent magnet fixed on a translation stage, allowing micrometric movement along the three
spatial axes, provided the external magnetic field applied to the diamond sample. An additioanl coil (not shown in Fig.1)  provided magnetic field modulation. The coil is aligned along the vertical z-direction, in this way the modulated additional field has the same projection along all the four possible orientation of the NV axis. 

The photoluminescence (PL) emission was spectrally filtered with a notch filter and a long-pass filter both centered at 650 nm, then collected and detected with two different acquisition systems. A 4\% fraction of the total PL intensity was sent to a single-photon detector (SPD). The signal from the SPD was used for the ODMR spectrum acquisition. The remaining 96\% fraction of the emitted PL intensity was collected by NA=0.25 objective (Olympus 10$\times$) and imaged onto a photodetector (Thorlabs DET 10A2). The signal from the photodiode was sent to the input channel of the LIA.

The time constant of the LIA was set to $\tau=10$ ms for the construction of the LIA spectrum. While scanning the microwave, 20 independent measurements of the LIA signal were acquired. In the measurement of the Linear Spectral Density(LSD) of the
noise, a time constant of $\tau=1$ ms was set. For the estimation of LSD, we acquired the LIA signal for 10 minutes with a sampling rate of $s=5000$ Hz and subsequently the LIA signal was Fast-Fourier transformed.

\section{Results}

\begin{figure}[!h]
\centering
\includegraphics[scale=0.5]{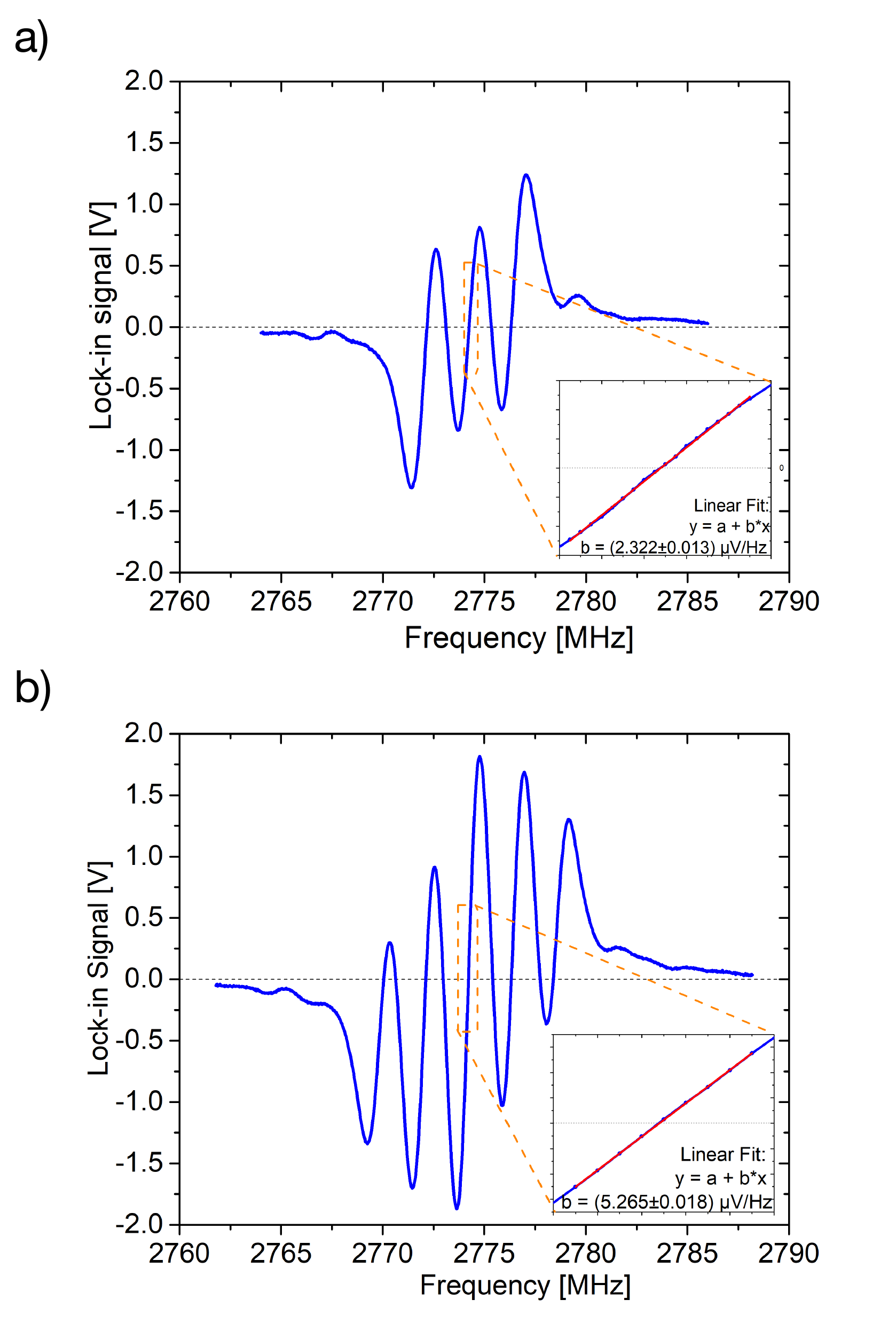}
\caption{ Lock-in spectrum with excitation of a single resonance (a) and simultaneous excitation of the three resonances separated by the hyperfine coupling (b). The insets shows the central linear zone and the value $b$ of the slope the curve.}
\label{fig:lock_in}
\end{figure}

\begin{figure}[!h]
\centering
\includegraphics[scale=0.3]{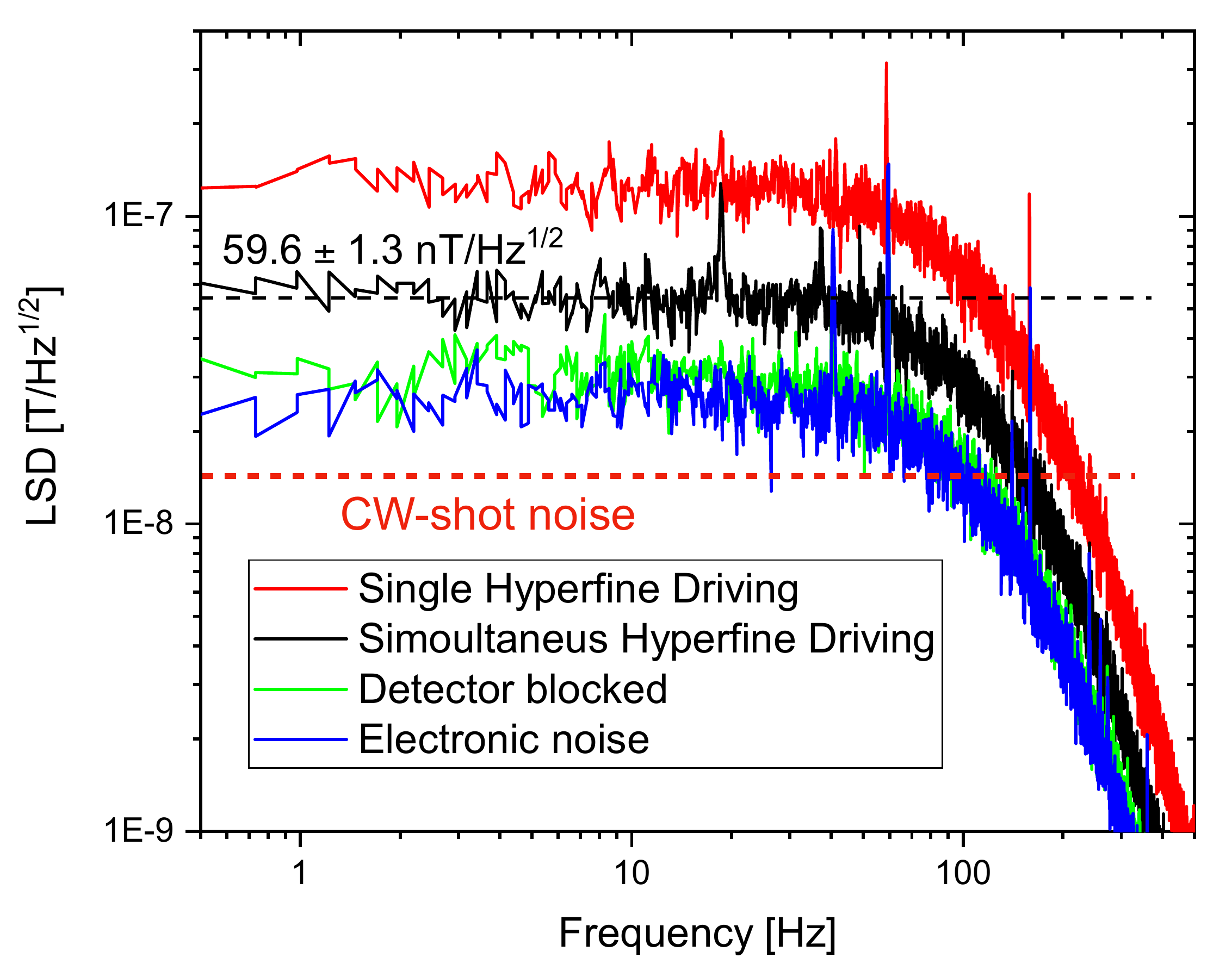}
\caption{ Comparison between the linear spectral densities (LSD) of the readout of the NV sensor in single hyperfine driving regime (red line) and in the simultaneous hyperfine driving (black line). The linear spectral density of the read-out with the detector blocked(green line), with the input of the LIA disconnected (blue line)  and the Continous Wave(CW)-shot noise (red dashed line) are also shown.}
\label{fig:LIA}
\end{figure}

\begin{figure}[!h]
\centering
\includegraphics[scale=0.3]{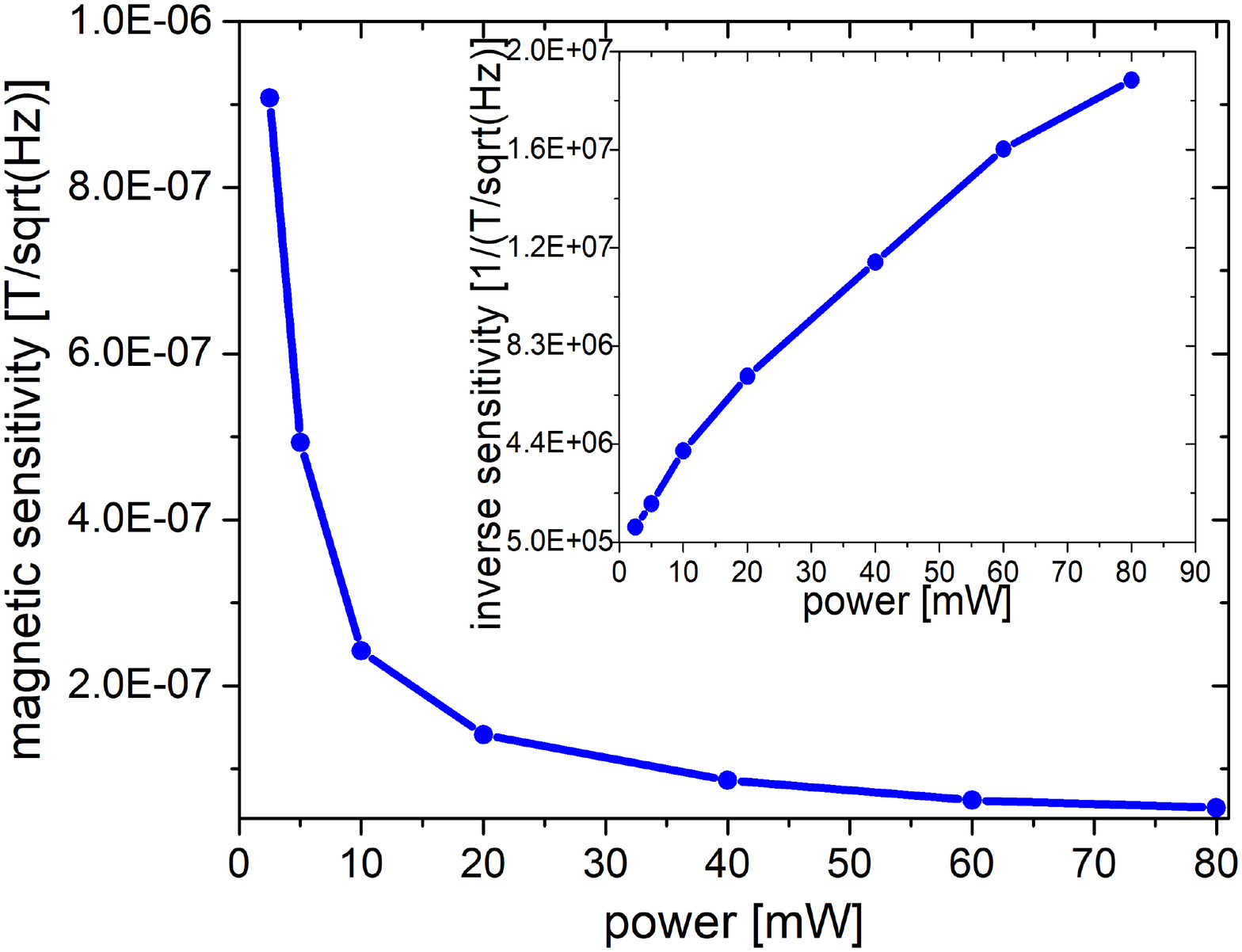}
\caption{Magnetic Sensitivity for different values of the applied optical power. In the inset, the inverse of the sensitivity is shown  for the same power values.}
\label{fig:SensVsPower}
\end{figure}

Figure \ref{fig:lock_in}(a) depicts the LIA signal in function of the microwave frequency. Three frequency ranges can be identified over which the LIA signal is directly proportional to the resonance shift and hence to the applied field. A yellow dashed rectangle encloses the central one.
These three zones correspond to the three dips in the ODMR spectrum due to the hyperfine coupling of the NV electronic system with the $^{14}$N nuclear spin. The LIA signal is linear in these zones because the LIA detection method is sensitive to the first derivative of the ODMR spectrum.

The figure of merit of the LIA detection method is represented by the slope $b$ of the curve in the linear zone, as reported in in Fig. \ref{fig:lock_in}(a). In this zone, $\delta B_{NV}$ is related to the measured LIA signal $S_{LIA}$ by\cite{barry2016optical} by:
\begin{equation}
\delta B_{NV}=\frac{h}{g\mu_B}\delta\nu_+=\frac{h}{g\mu_B}\frac{1}{b}S_{LIA}. 
\label{eq:LIA}
\end{equation}
It is possible to increase the slope of the curve (and thus the
sensitivity of the technique) by simultaneously addressing all the three resonances \cite{barry2016optical,el2017optimised}. To this scope, three frequency-modulated microwave tones separated by the hyperfine splitting $A_{orth}=2.16 $ MHz are generated. When the center tone is at the frequency of the center resonance, all three resonance are excited, thus enhancing the slope of the curve. Fig. \ref{fig:lock_in}(b) shows an example of the LIA  spectrum for multiple frequency excitation. $b$ is increased by a factor $\approx2$ compared to single-tone excitation. 

The minimum detectable field $B_{min}$ is
\begin{equation}\label{eq:eta}
 B_{min}= \frac{h}{g\mu_B}\frac{1}{b}\frac{\sigma_{S}}{\sqrt{N}},
\end{equation}
where we have considered $N$ independent measurements and that $S_{LIA}$ is affected by an uncertainty equal to $\sigma_{S}$. Increasing the total time of measurement $T$ leads to the usual scaling of the sensitivity $\eta$:
\begin{equation}\label{eq:eta_bis}
\eta = B_{min}\sqrt{T}=\frac{h}{g\mu_B}\frac{1}{b}\frac{\sigma_{S}}{\sqrt{N}}\sqrt{T},
\end{equation}

Fig. \ref{fig:LIA} shows the Linear Spectral Density of the LIA noise multiplied by the factor $\frac{g\mu_B\Delta\nu}{b}$, as defined in equation Eq. \ref{eq:eta}. For the Lock-in detection scheme described in this work, $\eta$ corresponds to the low-frequency plateau in Fig. \ref{fig:LIA}. 
Fig. \ref{fig:LIA} also shows the shot-noise limit for the single hyperfine driving. 
The Continuos Wave(CW) shot-noise limit is:
\begin{equation}\label{eq:sensitivity}
\eta_{CW}=\frac{h}{g\mu_B}\frac{\sqrt{I_0}}{\max\left(\frac{\partial I_0}{\partial\nu}\right)}
\end{equation}
We estimated it from ODMR spectrum as follow:
\begin{equation}\label{eq:sensitivity_bis}
\eta_{CW}=K\frac{h}{g\mu_B}\frac{\Delta\nu}{2\sqrt{I_0}C}=14 \,\, \text{nT/Hz}^{1/2}
\end{equation}
where $\Delta\nu=2.599 $ MHz is the linewidth, $C=0.006$ the contrast for the central dip of the hyperfine spectrum and $K=0.31$ is a specific constant of this line. $I_0=3.03\cdot10^{10}$ s$^{-1}$ is estimated from the optical power incident onto the photodiode $W=8.5$ nW, considering a photon energy $E_{ph}=2.84\cdot10^{-19}$ J.

Fig. 3 shows that simultaneous driving improves the sensitivity by a factor of $\approx$ 2. This improvement is due to two contributions: (i) an increase in the slope $b$ by a factor $\approx2$ (ii) no significant increase in the LIA noise. 

To point out the biocompatibility of this method, we measured the magnetic sensitivity for different applied powers, see Fig. \ref{fig:SensVsPower}. The sensitivity decreased by lowering the laser power.There is a tradeoff between laser power reduction that improves biocompatibility and sensitivity that will be discussed in detail in the next section. In the next section, we will discuss which maximum power a biological system can bear.

\section{Discussion}

\begin{figure}[!h]
\centering
\includegraphics[scale=0.3]{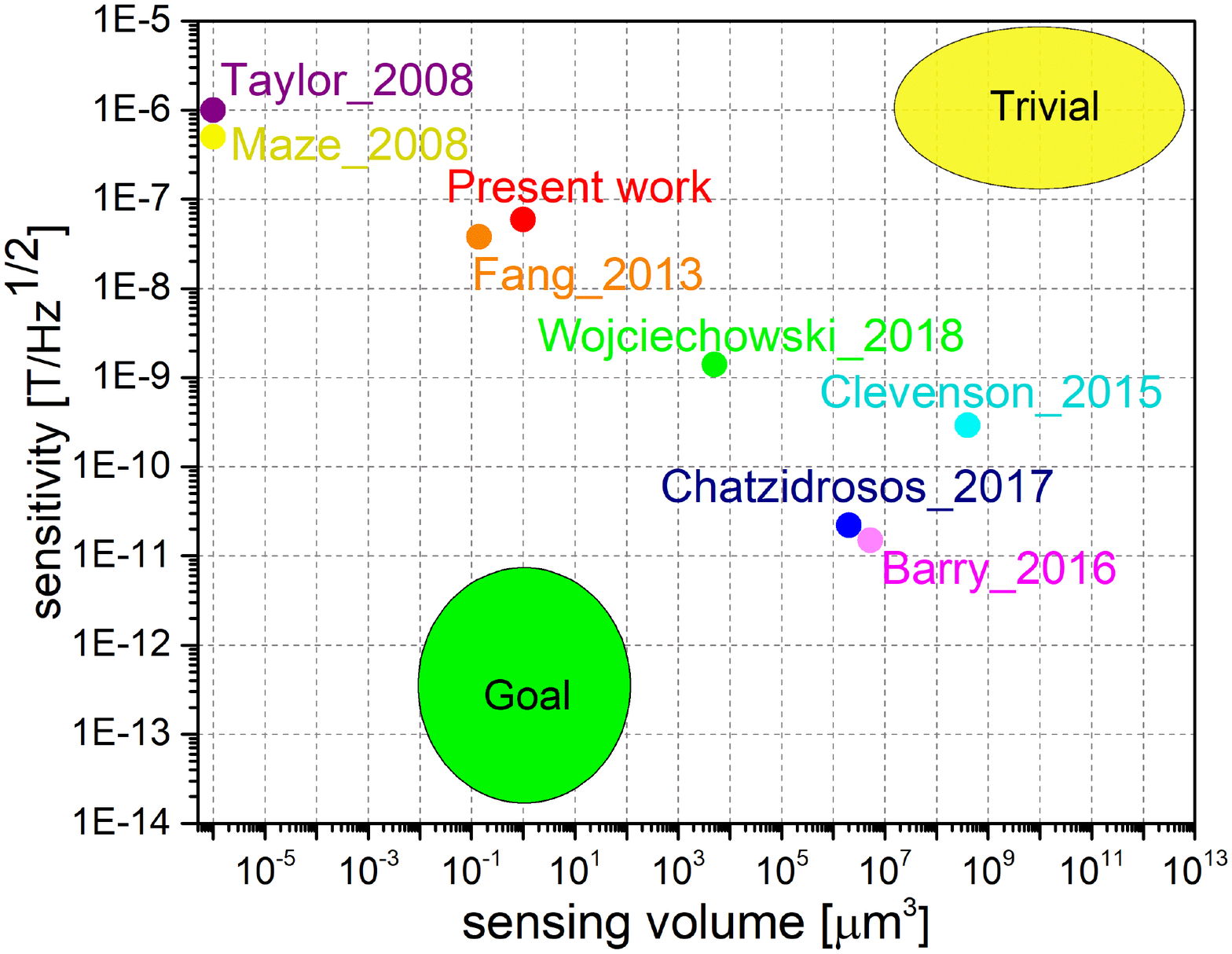}
\caption{The sensitivity in function of the sensing volume for the present study and  from data taken from literature[NV\cite{taylor2008high}, ND 2008\cite{maze2008nanoscale}, Fang 2013\cite{fang2013high}, Barry 2016\cite{barry2016optical}, Wojciechowski 2018 \cite{wojciechowski2018precision}, Clevenson 2015 \cite{clevenson2015broadband} ,  Chatzidrosos 2017 \cite{chatzidrosos2017miniature}]. The region of interest for biological application is defined by the green region}
\label{fig:sens_sens_vol}
\end{figure}

Here presented method is in principle able to resolve the contribution of a single cell to the magnetic field considering the fact that in our demonstrated case the sensing volume is defined by the laser spot in x-y plane of $(10 \times 10 )\,\mu\textrm{m}^2$), and the thickness of NV-rich layer of 10 nm (along z-axis).

We obtained a sensitivity of $\eta=59.6\pm1.3 \, \textrm{nT}/\sqrt{\textrm{Hz}}$ for an optical power of 80 mW: this value is beyond (or at least well comparable) with the one obtained in previous works if the sensing volume is taken in account \cite{barry2016optical, wojciechowski2018precision, clevenson2015broadband, chatzidrosos2017miniature}, see Fig.\ref{fig:sens_sens_vol}. Nonetheless, it has to be underlined that it is not proved that living cells can sustain 80 mW of power radiating on a surface of 100 $\mu$m$^2$, even considering that we apply this power only for a measurement time of 10 ms in a measurement cycle of 1 s. On the other end, living neuronal cells can surely tolerate without being affected a power of 3 mW applied for minutes in the same optical geometry of our setup\cite{guarina2018nanodiamonds}. Considering that we apply the optical power only for few milliseconds, we can estimate  a conservative biocompatible optical power $\geq10$ mW, that results in a sensitivity around  $\eta_{bio}\leq200\, \textrm{nT}/\sqrt{\textrm{Hz}}$.

This value of sensitivity still needs to be increased to sense neuron (or hearth cells) activity, where we expect a 1-10 nT in proximity of a single channel (a functionalised nanodiamond can in principle be targeted at nanometric distance from the channel) or when considering tissue slices. Furthermore, ion channels clustering can further increase the previous values.  

This is a reasonable improvement with present techniques: Barry et al. \cite{barry2016optical} estimated that a 300-fold improvement could be achieved using pulse sequences and new techniques in sample preparation. The main advantages offered by pulsed techniques are \cite{barry2020sensitivity}: (i) the possibility of working in a high-optical-intensity regime (ii) the possibility of taking advantage of a elongated coherence time $T^*_2$.  Also considering this improvement, a factor five  is still needed to achieve the magnetic sense of the heart activity. The use of a heat-sink and of a reflective layer could allow the adoption of higher optical powers and thus reach the desired sensitivity.

\section{Conclusions}

We presented an experimental apparatus and sensing protocol compatible with the measurement of magnetic fields in biological systems at an intracellular/cellular scale, and with a sensitivity beyond previous works. These
results indicate a clear strategy for magnetic sensing at cellular level, contributing to paving the way to practical biological applications of these methods.
Neverthless, it must be emphasized that due to the small, potentially at nanoscale, volume of our technique, this can find a broad application, beyond the biological example that we have discussed

\section*{Availability of data and materials}
All the experimental data presented are available from the authors upon reasonable request.


\section*{Funding}
This work has received funding from the European Union's PATHOS EU H2020 FET-OPEN grant no. 828946 and Horizon 2020, from the EMPIR Participating States in the context of the projects EMPIR-17FUN06 ''SIQUST'' and from the project Piemonte Quantum Enabling Technologies (PiQuET) funded by the Piemonte Region.



\section*{Acknowledgements}
The authors wish to thank Elio Bertacco for the technical help in implementing the frequency modulation and Giulia Tomagra for the interesting discussions about the biocompatibility of the setup.


\end{document}